\documentclass[aps,twocolumn,prd,amsmath,footinbib,preprintnumbers,superscriptaddress]{revtex4}
\usepackage{graphicx}
\usepackage{dcolumn}
\usepackage{bm}
\usepackage{xcolor}
\usepackage{amsmath}
\usepackage{amssymb}
\usepackage{booktabs,multirow,tabularx}
\usepackage{subfig}
\usepackage{comment}
\usepackage[colorlinks]{hyperref}
\hypersetup{colorlinks=true, linkcolor=blue, filecolor=blue, urlcolor=blue, citecolor=blue}

\newcommand{\ttbar}{\ensuremath{t\overline{t}}}

\newcommand{\etat}{\ensuremath{\eta_t}}
\newcommand{\Thetat}{\ensuremath{\Theta_t}}
\newcommand{\chit}{\ensuremath{\chi_{tJ}}}
\newcommand{\chitz}{\ensuremath{\chi_{t0}}}
\newcommand{\chitt}{\ensuremath{\chi_{t2}}}
\newcommand{\history}{\texttt{history}}
\newcommand{\madgraph}{\texttt{MadGraph5\_aMC@NLO}}

\newcommand{\madsons}{\texttt{MadSONS}}

\preprint{TTK-26-41, P3H-26-089}

\begin{document}

\title{Toponium hadroproduction at next-to-next-to-leading order in NRQCD}

\author{Sven Yannick Klein}
\affiliation{Institute for Theoretical Particle Physics and Cosmology, RWTH Aachen University, Sommerfeldstra{\ss}e 16, 52074 Aachen, Germany}

\author{Fabio Maltoni}
\affiliation{Centre for Cosmology, Particle Physics and Phenomenology (CP3), Universit\'e Catholique de Louvain, Chemin du Cyclotron, Louvain-la-Neuve, B-1348, Belgium}
\affiliation{Dipartimento di Fisica e Astronomia, Universit\`a di Bologna and INFN, Sezione di Bologna,
Via Irnerio 46, 40126 Bologna, Italy}

\author{Hua-Sheng Shao}
\affiliation{
 Laboratoire de Physique Th\'eorique et Hautes Energies (LPTHE), UMR 7589, Sorbonne Universit\'e et CNRS, 4 place Jussieu, 75252 Paris Cedex 05, France
}

\author{Lukas Simon}
\affiliation{
 Laboratoire de Physique Th\'eorique et Hautes Energies (LPTHE), UMR 7589, Sorbonne Universit\'e et CNRS, 4 place Jussieu, 75252 Paris Cedex 05, France
}
\affiliation{Dipartimento di Fisica, Università di Torino, and INFN, Sezione di Torino, Via P. Giuria 1, I-10125 Torino, Italy}

\author{Guoxing Wang}
\affiliation{
 Laboratoire de Physique Th\'eorique et Hautes Energies (LPTHE), UMR 7589, Sorbonne Universit\'e et CNRS, 4 place Jussieu, 75252 Paris Cedex 05, France
}
\affiliation{School of Physics, East China University of Science and Technology, 130 Meilong Road, Shanghai, 200237, China}

\date{\today}

\begin{abstract}
We present the first next-to-next-to-leading-order (NNLO)
predictions for inclusive color-singlet toponium hadroproduction
within  NRQCD in the zero-width limit for the top
quark. The central new result is the NNLO QCD correction to the
short-distance coefficient for pseudoscalar toponium production.
The predictions include electroweak and relativistic corrections,
additional production channels, and perturbatively evaluated
bound-state matrix elements. At the Large Hadron Collider proton-proton center-of-mass
energy of $13\,\mathrm{TeV}$, we obtain a cross section of
$4.87^{+0.63}_{-0.65}\,\mathrm{pb}$, with scale and parametric
uncertainties combined. The NNLO corrections reduce the
renormalization- and factorization-scale uncertainties by
approximately a factor of four relative to NLO, to the
few-percent level. The dominant remaining uncertainty arises
from the Coulomb-scale dependence of the bound-state matrix
elements.
\end{abstract}

\maketitle

\textit{Introduction} -- Top-quark pair production near threshold probes QCD binding in
a regime accessible to perturbation theory. In the attractive
color-singlet channel, repeated Coulomb exchange generates
would-be bound-state poles, which the finite top-quark width
smears into a broad enhancement of the production
rate~\cite{Fadin:1987wz,Fadin:1990wx}.

These dynamics have long been studied in $e^+e^-$ annihilation,
where the dominant contribution near threshold is the
color-singlet, spin-triplet $S$-wave channel $\Thetat$.
QCD predictions have reached next-to-next-to-next-to-leading
order (N$^3$LO)~\cite{Beneke:2015kwa},
and the resummation of velocity logarithms has been developed
at next-to-next-to-leading logarithmic (NNLL)
accuracy~\cite{Hoang:2013uda}. This theoretical framework
underpins the proposed FCC-ee threshold scan, which aims to
determine a short-distance top-quark mass with an experimental
precision of order $20\,\mathrm{MeV}$ or
better~\cite{FCC:2018byv,Defranchis:2025auz}.

At hadron colliders, several spin and color configurations
contribute near threshold. The leading bound-state enhancement
arises from the color-singlet, spin-singlet $S$-wave state
$\etat$~\cite{Hagiwara:2008df,Kiyo:2008bv}.
Its characteristic spin correlations provide a means of
identifying toponium effects beyond an enhancement of the
production rate~\cite{Fuks:2021xje}. Such spin information also
underlies studies of quantum observables in top-quark
pairs~\cite{Afik:2020onf,Fabbrichesi:2021npl,Severi:2021cnj}.
At the Large Hadron Collider (LHC), CMS and ATLAS have now observed an excess of events near the
$\ttbar$ threshold in the dilepton channel, consistent with a
color-singlet pseudoscalar
contribution~\cite{CMS:2025kzt,ATLAS:2026dbe}.
CMS has independently observed a compatible excess in the
single-lepton channel~\cite{CMS:2026ani}.

Precise Standard Model predictions are needed to quantify the
threshold enhancement relative to calculations without
resummation and to assess any remaining discrepancy with the
data. This requires control of both hard production and
nonrelativistic dynamics. 
Joint resummation of soft-gluon and Coulomb effects,
including bound-state contributions, has been developed
for inclusive hadronic top-quark pair
production~\cite{Beneke:2011mq}.
Calculations have progressively refined the threshold spectrum, its theoretical uncertainties, and its
implementation in event
generators~\cite{Sumino:2010bv,Ju:2020otc,Fuks:2024yjj,Garzelli:2024uhe,Nason:2025hix,Garzelli:2026ctb,Sjostrand:2026rcz,Nason:2026oka}.
For pseudoscalar color-singlet hadroproduction, the
next-to-leading-order (NLO) QCD corrections to the
short-distance coefficient have long been
known~\cite{Kuhn:1992qw,Petrelli:1997ge}, but the corresponding
complete next-to-next-to-leading-order (NNLO) QCD correction
has remained missing. Determining this correction is essential
to establish the perturbative stability of the production rate
and to disentangle uncertainties in hard production from those
associated with binding dynamics.

In this Letter, we compute the NNLO QCD correction to the
short-distance coefficient for $pp\to\etat+X$. We use this
result to obtain NNLO predictions for inclusive color-singlet
toponium hadroproduction within nonrelativistic
QCD (NRQCD)~\cite{Bodwin:1994jh}, working in the zero-width limit for
the top quark. The predictions include electroweak (EW) and
relativistic corrections, the additional production channels
required by our power counting, and perturbatively evaluated
bound-state matrix elements.
The new short-distance coefficients provide a first necessary step towards an NNLO prediction for the physical top-quark pair threshold enhancement; further details of the present calculation and its extension to a Green’s-function treatment incorporating the finite top-quark width will be presented in a forthcoming publication.

\textit{Methodology} -- Within NRQCD,
we adopt the power counting
$\alpha_s\sim v\sim\sqrt{\alpha}\ll1$, where $v$ denotes the
nonrelativistic top-quark velocity. We treat the top-quark mass $m_t$, the $W$-boson mass $m_W$,
the $Z$-boson mass $m_Z$, and the Higgs-boson mass $m_H$ as hard scales, with
$m_{t,W,Z,H}\gg m_tv\gg m_tv^2$, and retain contributions
through relative order $\mathcal{O}(\alpha_s^2)$ to
color-singlet toponium hadroproduction. The cross section
factorizes into short-distance coefficients (SDCs) and
long-distance matrix elements (LDMEs). Through NNLO, it reads
\begin{eqnarray}
\sigma_{\rm NNLO}&=&
\hat{\sigma}_{\etat}\langle\mathcal{O}_{\etat}\rangle
+\hat{\sigma}_{\Thetat}\langle\mathcal{O}_{\Thetat}\rangle\nonumber\\
&+&\hat{\sigma}_{\etat,v^2}\langle\mathcal{P}_{\etat}\rangle
+\hat{\sigma}_{\chitz}\langle\mathcal{O}_{\chitz}\rangle
+\hat{\sigma}_{\chitt}\langle\mathcal{O}_{\chitt}\rangle\,,
\label{eq:sigmadecom}
\end{eqnarray}
where $\etat$ and $\Thetat$ denote the spin-singlet and
spin-triplet $S$-wave states, respectively, and $\chit$
denotes the spin-triplet $P$-wave states with $J=0,2$.
The leading contribution arises from $gg\to\etat$:
the leading order (LO) SDC $\hat{\sigma}_{\etat}^{(0)}$ scales as
$\alpha_s^2$, while the LO LDME scales as $\alpha_s^3$,
giving an overall cross section of order $\alpha_s^5$.
The third term in eq.~\eqref{eq:sigmadecom} describes the
relativistic $\mathcal{O}(v^2)$ correction to $\etat$ production,
and the last two terms account for the $P$-wave contributions.
Explicit expressions for the LDMEs
are collected in the Supplemental Material~\ref{sec:SMLDME}.

\textit{Short-distance coefficients} -- For pseudoscalar
production, the NNLO QCD correction
$\hat{\sigma}_{\etat}^{(2)}$ to the SDC constitutes our main
new result. The corresponding NLO correction,
$\hat{\sigma}_{\etat}^{(1)}$, has long been
known~\cite{Kuhn:1992qw,Petrelli:1997ge}. We perform the NNLO calculation
with \history~\cite{Klein:2026tlj}, using nested
soft-collinear subtraction~\cite{Caola:2017dug,Caola:2019nzf}
and the two-loop amplitudes of
refs.~\cite{Abreu:2022vei,Abreu:2022cco}. To improve
perturbative stability, we use the soft-improved matching
(SIM) scheme~\cite{Maxia:2026fbz} for the NRQCD matching.

To complete $\hat{\sigma}_{\etat}$ through NNLO in our power
counting, we also include the NLO EW correction
$\hat{\sigma}_{\etat}^{({\rm EW})}$ to $gg\to\etat$ at
$\mathcal{O}(\alpha_s^2\alpha)\sim\mathcal{O}(\alpha_s^4)$,
the tree-level associated-production processes
$gg\to\etat H$ and $gg\to\etat Z$ at
$\mathcal{O}(\alpha_s^2\alpha)$, and
$pp\to\etat\ttbar$ at $\mathcal{O}(\alpha_s^4)$. We collectively denote these associated-production contributions by $\hat{\sigma}_{\etat}^{(\mathrm{assoc.})}$.

For the spin-triplet state, charge-conjugation (C) parity
forbids $gg\to\Thetat$, so $\hat{\sigma}_{\Thetat}^{(0)}$ vanishes
at $\mathcal{O}(\alpha_s^2)$. The leading QCD contribution,
$\hat{\sigma}_{\Thetat}^{(1)}$, arises from
$gg\to\Thetat g$ at $\mathcal{O}(\alpha_s^3)$, while the NLO QCD correction to this process, $\hat{\sigma}_{\Thetat}^{(2)}$,
enters at $\mathcal{O}(\alpha_s^4)$. These contributions
therefore enter the total production rate at NLO and NNLO,
respectively. The NLO QCD correction to $gg\to\Thetat g$ has long been known for
the analogous charmonium and bottomonium
processes~\cite{Campbell:2007ws}. We implement it in
\history\ and validate the result against an independent
implementation based on \madgraph~\cite{Alwall:2014hca}.

The tree-level EW processes
$q\overline{q}\to\gamma^*/Z^*\to\Thetat$ contribute at
$\mathcal{O}(\alpha^2)\sim\mathcal{O}(\alpha_s^4)$
and thus also enter the total rate at NNLO. The corresponding SDC is denoted by $\hat{\sigma}_{\Thetat}^{(q\bar{q}\to\gamma^*/Z^*\to\Thetat)}$.
We additionally include the tree-level associated-production
processes $gg\to\Thetat\gamma$, $gg\to\Thetat Z$, and
$pp\to\Thetat\ttbar$, which contribute at the same perturbative order. Their combined SDC contribution is denoted by $\hat{\sigma}_{\Thetat}^{(\mathrm{assoc.})}$. These associated production processes involving $\etat$ and $\Thetat$, as well as the process $q\overline{q}\to\gamma^\ast/Z^\ast\to\Thetat$, can be simulated using the \madsons\ module~\cite{ColpaniSerri:2025vdz,Maxia:2026ved} in \madgraph. The tree-level process $gg\to\Thetat H$ is forbidden by C-parity conservation~\footnote{The contribution $q\overline{q}\to\gamma^*/Z^*\to\etat$ vanishes due to C-parity conservation and the chirality flip for massless initial-state quarks. We also do not consider $\gamma\gamma\to \etat$, whose SDC formally contributes at NNLO, at $\mathcal{O}(\alpha^2)\sim\mathcal{O}(\alpha_s^4)$. However, this contribution is further suppressed by the photon PDF, which is of $\mathcal{O}(\alpha)$ relative to the corresponding quark or gluon PDFs. The study of this process in ultraperipheral collisions at hadron colliders can be found in ref.~\cite{dEnterria:2025ecx}.}.

In this work, we compute the inclusive rate through NNLO in our power counting, including all associated-production channels. Their combined contribution is very small, but these channels may be experimentally distinguishable by tagging the associated particles; we therefore report their individual rates in the Supplemental Material \ref{sec:SMtables}.

\textit{Long-distance matrix elements} -- We evaluate the
$S$-wave LDMEs $\langle\mathcal{O}_{\etat}\rangle$ and
$\langle\mathcal{O}_{\Thetat}\rangle$ including NNLO QCD
corrections~\cite{Melnikov:1998ug,Beneke:1999qg,Kniehl:2002yv,Penin:2004ay,Beneke:2005hg,Kiyo:2010jm}
and NLO QED corrections~\cite{Pineda:2006ri}.
For each spin channel, the LDME includes a sum over all radial
bound states, as specified in the Supplemental Material \ref{sec:SMLDME}.
The QCD contributions are evaluated in the SIM scheme and
further improved by renormalization-group (RG) evolution. 

We also include the leading top-Yukawa
contribution~\cite{Beneke:2015lwa}. Although formally of
N$^3$LO in our power counting, this term is retained because
of the large coupling $y_t\simeq1$.
For the relativistic and $P$-wave contributions, which first
enter the total rate at NNLO, the LO expressions for
$\langle\mathcal{P}_{\etat}\rangle$,
$\langle\mathcal{O}_{\chitz}\rangle$, and
$\langle\mathcal{O}_{\chitt}\rangle$ suffice (see the Supplemental Material~\ref{sec:SMLDME}).

The LDMEs depend on the Coulomb (soft) scale $\mu_C$,
which we choose to be of the order of a reference Bohr
scale $\mu_B=a_B^{-1}$, defined implicitly by
\begin{equation}
\mu_B=e^{\gamma_E}C_F\alpha_s(\mu_B)m_t\,,
\end{equation}
where $C_F=4/3$ is the quadratic Casimir of the fundamental
representation and $\gamma_E$ is the Euler--Mascheroni
constant. Our scale convention includes the factor
$e^{\gamma_E}\simeq1.78$, also used in
ref.~\cite{AbdulHameed:2026utd}, motivated by the form of the
zero-distance Coulomb Green's function in the threshold
limit~\cite{Fadin:1987wz}.

\textit{Results} -- We use the
\texttt{NNPDF40\_nnlo\_as\_01180} parton distribution functions (PDF)
set~\cite{NNPDF:2021njg}, accessed through
\texttt{LHAPDF6}~\cite{Buckley:2014ana}, together with its
associated running of $\alpha_s$. All quarks except the top
are treated as massless. Following the 2026 Particle Data 
Group~\cite{ParticleDataGroup:2026aaa}, we take
$m_t=172.6\,\mathrm{GeV}$,
$m_H=125.13\,\mathrm{GeV}$,
$m_W=80.3625\,\mathrm{GeV}$, and
$m_Z=91.1879\,\mathrm{GeV}$.
We adopt the $G_\mu$ EW scheme with
$G_F=1.1663785\times10^{-5}\,\mathrm{GeV}^{-2}$,
corresponding to $\alpha\simeq1/132.047$.

Our central renormalization and factorization scales are
$\mu_R=\mu_F=m_t$. We estimate their uncertainty through
the standard seven-point variation,
$\mu_R,\mu_F\in\{m_t/2,m_t,2m_t\}$, subject to
$1/2\leq\mu_R/\mu_F\leq2$.
For the LDMEs, we choose $\mu_C=\mu_B$ centrally and
estimate the Coulomb-scale uncertainty using
$\mu_C=\xi_C\mu_B$ with $\xi_C\in\{1/2,1,2\}$.
To isolate the perturbative corrections to the SDCs, we use
the same LDME values at LO, NLO, and NNLO. These order labels
therefore specify the short-distance contributions retained
according to our power counting. At LO, the cross section is given by
\begin{equation}
\sigma_{\rm LO}
=\hat{\sigma}_{\etat}^{(0)}
 \langle\mathcal{O}_{\etat}\rangle\,,
\end{equation}
where $\hat{\sigma}_{\Thetat}^{(0)}=0$, and the LO SDC  for the $\etat$ state is
\begin{equation}
\hat{\sigma}_{\etat}^{(0)}=\frac{\pi^3
\alpha_s^2(\mu_R)}{144m_t^5}\frac{4m_t^2}{s}\mathcal{L}_{gg}\!\left(\frac{4m_t^2}{s},\mu_F\right)\,,
\end{equation}
where $\mathcal{L}_{gg}$ is the gluon-gluon 
partonic luminosity.
The NLO prediction is then given by
\begin{equation}
\begin{aligned}
\sigma_{\rm NLO}={}&
\left(\hat{\sigma}_{\etat}^{(0)}
+\hat{\sigma}_{\etat}^{(1)}\right)
\langle\mathcal{O}_{\etat}\rangle+\hat{\sigma}_{\Thetat}^{(1)}
\langle\mathcal{O}_{\Thetat}\rangle\,.
\end{aligned}
\end{equation}
The NNLO prediction is obtained from
eq.~\eqref{eq:sigmadecom}, with
\begin{eqnarray}
\hat{\sigma}_{\etat}&=&\sum_{i=0}^{2}{\hat{\sigma}_{\etat}^{(i)}}+\hat{\sigma}_{\etat}^{(\mathrm{EW})}+\hat{\sigma}_{\etat}^{(\mathrm{assoc.})}\,,\\
\hat{\sigma}_{\Thetat}&=&\hat{\sigma}_{\Thetat}^{(1)}+\hat{\sigma}_{\Thetat}^{(2)}+\hat{\sigma}^{(q\bar{q}\to\gamma^*/Z^*\to \Thetat)}_{\Thetat}+\hat{\sigma}_{\Thetat}^{(\mathrm{assoc.})}\,.
\end{eqnarray}
The remaining SDCs, evaluated at $\mathcal{O}(\alpha_s^2)$, are
\begin{equation}
\hat{\sigma}_{\etat,v^2}=-\frac{4}{3m_t^2}\hat{\sigma}_{\etat}^{(0)}\,,
\hat{\sigma}_{\chitz}=\frac{3}{m_t^2}\hat{\sigma}_{\etat}^{(0)}\,,
\hat{\sigma}_{\chitt}=\frac{4}{5m_t^2}\hat{\sigma}_{\etat}^{(0)}\,.
\end{equation}

\begin{figure}
    \centering
    \includegraphics[height=8.6cm]{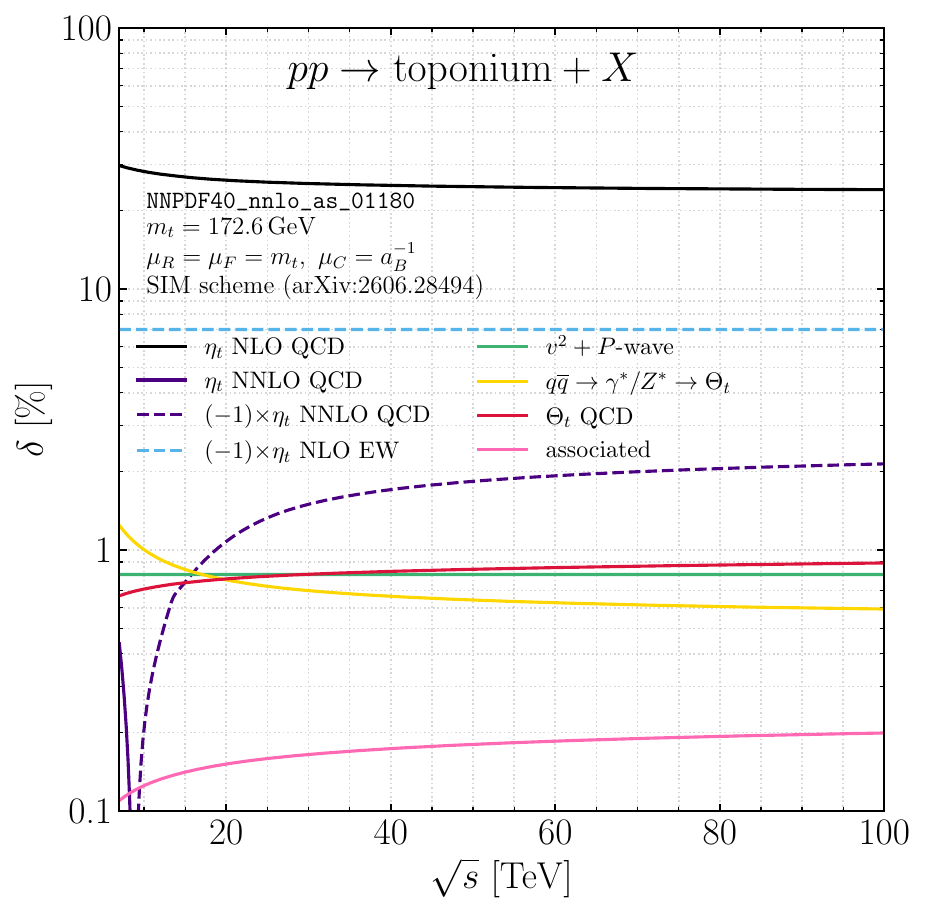}
    \caption{Relative corrections to the LO cross section as functions of $\sqrt{s}$. The dashed curves have been multiplied by $-1$ for better visibility.}
    \label{fig:xsec_decomposition}
\end{figure}

Figure~\ref{fig:xsec_decomposition} shows the corrections
relative to $\sigma_{\rm LO}$ as functions of $\sqrt{s}$
from $7$ to $100\,\mathrm{TeV}$. For the reader's convenience, the corresponding results are also provided in table~\ref{tab:xsec_decomposition} in the Supplemental Material \ref{sec:SMtables}. The largest contribution
is the NLO QCD correction to $\etat$ production, which
increases the LO cross section by approximately
$24\%$--$30\%$. The NLO EW correction reduces it by
approximately $7\%$, with no dependence on $\sqrt{s}$.
C parity forbids the real-emission process
$gg\to\etat\gamma$, so this EW correction follows the
gluon-luminosity dependence of the Born process. Its sizable magnitude is driven by the large Higgs-exchange contribution proportional to $y_t^2$, which dominates the EW correction.

The NNLO QCD correction is considerably smaller. It
changes sign near $\sqrt{s}\simeq9\,\mathrm{TeV}$ and reduces
the LO cross section by approximately $1\%$ at
$18\,\mathrm{TeV}$ and $2\%$ at $100\,\mathrm{TeV}$.
The combined relativistic and $P$-wave correction, the
$\Thetat$ QCD contribution ($\hat{\sigma}_{\Thetat}^{(1)}+\hat{\sigma}_{\Thetat}^{(2)}$), and the sum of the
associated-production contributions each remain below
$1\%$ in magnitude. The EW channel
$q\overline{q}\to\gamma^*/Z^*\to\Thetat$ exceeds this level
only for $\sqrt{s}\lesssim10\,\mathrm{TeV}$.
The small $\Thetat$ QCD contribution and the small associated-production rates
can be understood qualitatively from phase-space
suppression and the reduced parton luminosities
associated with additional hard particles in the
final state.

Associated-production channels contribute only at the per-mille level relative to $\sigma_{\mathrm{LO}}$ over the energy range considered. Channels with an associated EW boson dominate, while those with an additional $t\bar t$ pair are substantially smaller. The breakdown is reported in the Supplemental Material \ref{sec:SMtables}.

\begin{figure}
    \centering
    \includegraphics[height=8.6cm]{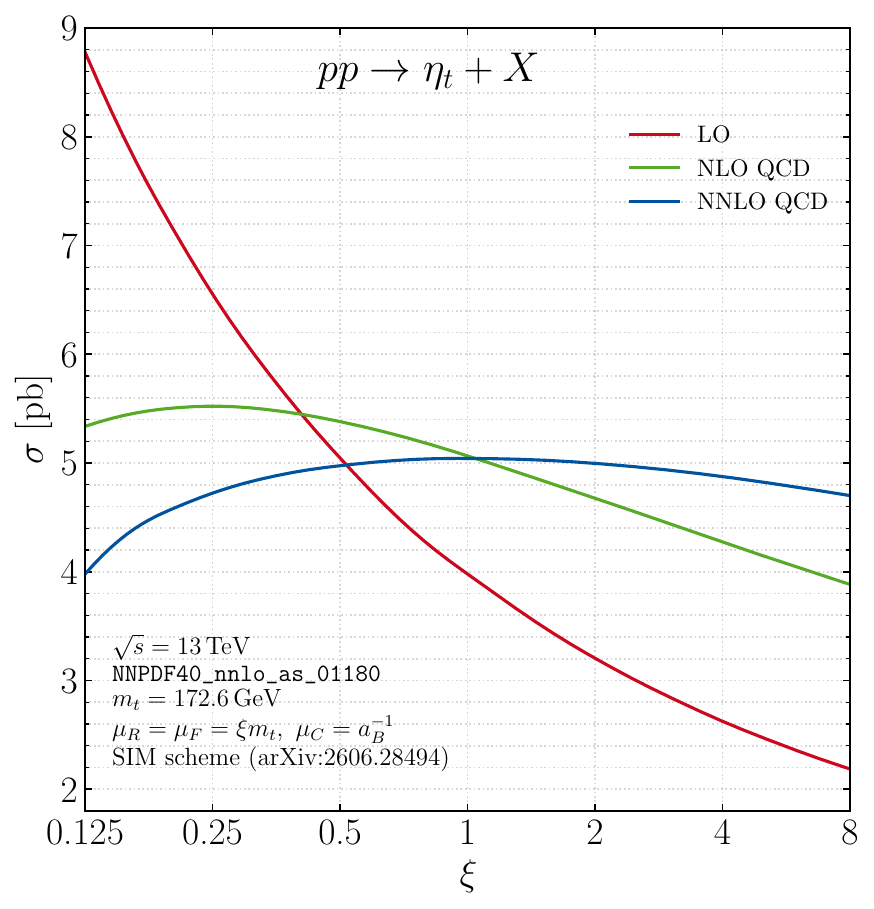}
    \caption{Renormalization- and factorization-scale dependence of the total LO, NLO QCD, and NNLO QCD cross sections for $pp\to \etat+X$ at the LHC with $\sqrt{s}=13\,\mathrm{TeV}$. We set $\mu_R=\mu_F=\xi m_t$.}
    \label{fig:scale_dependence}
\end{figure}
The perturbative stability of the prediction is illustrated in
figure~\ref{fig:scale_dependence}, which shows the LO, NLO QCD,
and NNLO QCD cross sections for $pp\to\etat+X$ at
$\sqrt{s}=13\,\mathrm{TeV}$ as functions of the common scale
$\mu_R=\mu_F=\xi m_t$, varied over $1/8<\xi<8$.
The scale dependence is substantially reduced at NLO and
further suppressed at NNLO, demonstrating improved perturbative
convergence. Interestingly, the small NNLO QCD correction at the
default central scale $\mu_R=\mu_F=m_t$ is not representative
of the full scale range: for scales above $2m_t$, the NNLO
QCD correction becomes significantly larger. This illustrates
that the apparent size of higher-order corrections can depend
strongly on the choice of the central scale and highlights the
importance of assessing perturbative stability through scale
variations rather than the size of individual corrections.

Table~\ref{tab:xsec_total} (see also figure~\ref{fig:xsec_total} in the Supplemental Material~\ref{sec:SMtables}) shows the total toponium-production cross sections at LO, NLO, and NNLO for LHC and FCC-hh energies, together with the uncertainties obtained from seven-point variations of $\mu_R$ and $\mu_F$. The $\mu_R$ dependence enters both the SDCs and the LDMEs but is dominated by the former. The increasing parton luminosities drive the NNLO cross
section from $1.17\,\mathrm{pb}$ at
$\sqrt{s}=7\,\mathrm{TeV}$ to $147\,\mathrm{pb}$ at
$100\,\mathrm{TeV}$. At $\sqrt{s}=13\,\mathrm{TeV}$,
the NLO corrections increase the LO prediction by
approximately $28\%$, while the NNLO corrections reduce
the NLO result by approximately $5\%$.
The width of the relative scale-variation decreases
by more than a factor of three at NLO and by a further
factor of approximately four at NNLO.
At $13\,\mathrm{TeV}$, the residual NNLO scale uncertainty
is $^{+0.5\%}_{-3.0\%}$, with a similar relative magnitude
throughout the energy range considered.

%=======================================================================%
% TABLE: Figure 3
%=======================================================================%
\begin{table}[t]

\centering
\renewcommand*{\arraystretch}{1.2}
\setlength{\tabcolsep}{9pt} % default: 6pt
\begin{tabular}[t]{|c|c|c|c|}
\hline
 $\sqrt{s}$ [TeV] & $\sigma_{\mathrm{LO}}$ $[\mathrm{pb}]$ & $\sigma_{\mathrm{NLO}}$ $[\mathrm{pb}]$ & $\sigma_{\mathrm{NNLO}}$ $[\mathrm{pb}]$ \\
\hline
$7$ & $0.93^{+0.31}_{-0.21}$ & $1.21^{+0.08}_{-0.12}$ & $1.17^{+0.01}_{-0.04}$ \\
$8$ & $1.30^{+0.42}_{-0.29}$ & $1.70^{+0.12}_{-0.16}$ & $1.63^{+0.01}_{-0.05}$ \\
$13$ & $3.98^{+1.07}_{-0.78}$ & $5.10^{+0.33}_{-0.40}$ & $4.87^{+0.03}_{-0.15}$ \\
$13.6$ & $4.38^{+1.15}_{-0.84}$ & $5.61^{+0.36}_{-0.43}$ & $5.34^{+0.03}_{-0.16}$ \\
$14$ & $4.65^{+1.21}_{-0.89}$ & $5.96^{+0.39}_{-0.45}$ & $5.67^{+0.03}_{-0.17}$ \\
$27$ & $16.4^{+3.7}_{-2.7}$ & $20.7^{+1.3}_{-1.4}$ & $19.6^{+0.1}_{-0.5}$ \\
$85$ & $100^{+23}_{-16}$ & $126^{+7}_{-8}$ & $118^{+1}_{-3}$ \\
$100$ & $126^{+29}_{-21}$ & $157^{+9}_{-11}$ & $147^{+1}_{-4}$ \\
\hline
\end{tabular}
\caption{Total inclusive toponium-production cross sections at LO, NLO, and NNLO as functions of $\sqrt{s}$. The quoted uncertainties arise solely from the standard seven-point variations of the renormalization and factorization scales.}
\label{tab:xsec_total}
\end{table}

The dominant remaining uncertainty arises from the
Coulomb-scale dependence of the LDMEs and amounts to
approximately $\pm12\%$. With common LDME inputs used at
all orders, this relative uncertainty is nearly independent
of the perturbative order and collision energy.
It could be reduced by extending the $S$-wave LDME
calculations to N$^3$LO; the corresponding QCD corrections
are already known for the spin-triplet
state~\cite{Beneke:2024sfa}.

The next-largest uncertainty comes from $\alpha_s(m_Z)$,
which enters both the SDCs and the LDMEs. Varying
$\alpha_s(m_Z)=0.118\pm0.001$ with the
\texttt{NNPDF40\_nnlo\_pdfas} set changes the cross section
by approximately $\pm5\%$. This dependence is driven
primarily by the LDMEs and is therefore also largely
independent of the perturbative order and $\sqrt{s}$.
The sensitivity to the top-quark mass is smaller:
varying $m_t=172.60\pm0.27\,\mathrm{GeV}$ changes the
cross section by approximately $\pm0.8\%$.
Combining these contributions with the renormalization-
and factorization-scale uncertainties in quadrature gives
\begin{equation}
\sigma_{\rm NNLO}
=4.87^{+0.63}_{-0.65}\,\mathrm{pb}
\qquad\text{at}\qquad
\sqrt{s}=13\,\mathrm{TeV}\,.
\end{equation}
This cross section quantifies the bound-state contribution to the top-quark pair threshold enhancement in our zero-width approximation, summed over all states strictly below threshold, $M_{\ttbar}<2m_t$.

\textit{Conclusion} -- We have presented the first NNLO
predictions for inclusive color-singlet toponium
hadroproduction within NRQCD, adopting the power counting
$\alpha_s\sim v\sim\sqrt{\alpha}$ and working in the
zero-width limit for the top quark. The central new result
is the NNLO QCD correction
to the short-distance
coefficient $\hat{\sigma}_{\etat}^{(2)}$ for $pp\to\etat+X$.
The NNLO corrections induce only a modest shift relative
to NLO and reduce the residual renormalization- and
factorization-scale dependence to the few-percent level.
The EW correction is sizable, reducing the LO
cross section by approximately $7\%$, and is dominated
by Higgs-exchange contributions proportional to $y_t^2$.
The dominant remaining uncertainty arises from
the Coulomb-scale dependence of the bound-state matrix
elements, making their higher-order evaluation a priority
for further improvement.

The new short-distance coefficients provide an essential ingredient for extending finite-width threshold calculations to NNLO within the Green’s-function approach~\cite{Fadin:1987wz,Kiyo:2008bv}. This extension will connect the improved production coefficients to the physical top-quark pair threshold spectrum measured at the LHC.

\textit{Acknowledgments} -- We thank S.~Abreu, M.~Becchetti, C.~Duhr, and M.~Ozcelik for providing us with the two-loop amplitudes presented in refs.~\cite{Abreu:2022vei,Abreu:2022cco}.
S.Y.K.\@ was supported by the Deutsche Forschungsgemeinschaft (DFG, German Research Foundation) under grant 396021762 -- \textit{TRR 257: P3H – Particle Physics Phenomenology after the Higgs Discovery} and received funding from the Federal Ministry of Education and Research (BMBF) under grant number 05H21PACCA.  F.M.\@ acknowledges support by the
F.R.S.–FNRS (Belgian Fund for Scientific Research) through the IISN convention 4.4517.08 -- \textit{Theory of fundamental interactions}.
H.-S.S., L.S., and G.W.\@ are supported by the grants from the ERC (grant 101041109 -- \textit{BOSON}) and the French ANR (grant ANR-20-CE31-0015 -- \textit{PrecisOnium}). Views and opinions expressed are however those of the authors only and do not necessarily reflect those of the European Union or the European Research Council Executive Agency. Neither the European Union nor the granting authority can be held responsible for them.
Computations were performed with computing resources granted by RWTH Aachen University under project rwth1826.

\bibliographystyle{utphys}

\bibliography{reference}

\onecolumngrid

\newpage
\appendix

\section{Supplemental Material}\label{sec:SM}

In this Supplemental Material, we present the NNLO expressions for the LDMEs in section~\ref{sec:SMLDME}, as well as the cross-section results in section~\ref{sec:SMtables}.

\subsection{Long-distance matrix elements\label{sec:SMLDME}}

The color-singlet $S$-wave LDME for the $n^{\mathrm{th}}$ radial excitation is related to the Schr\"odinger wavefunction at the origin through
\begin{equation}\label{eq:top_ldme}
	\left\langle\mathcal{O}^{\rm SIM}_{n,s}\right\rangle
	=(2s+1)\dfrac{2N_c}{4\pi}\left[\big\vert R^{\rm SIM}_{n,s}(0)\big\vert^2+\big\vert R^{\rm QED}_{n}(0)\big\vert^2+\big\vert R^{\rm Yuk}_{n}(0)\big\vert^2\right],
\end{equation}
where $N_c=3$, $s\in\{0,1\}$ is the spin quantum number, and $n\in\mathbb{N}$ is the principal quantum number. Owing to the large top-quark mass, the wavefunction at the origin can be calculated perturbatively. The NNLO QCD result in the $\overline{\rm MS}$ scheme is known from refs.~\cite{Melnikov:1998ug,Beneke:1999qg,Kniehl:2002yv,Penin:2004ay,Beneke:2005hg,Kiyo:2010jm}. The renormalization-group-improved NNLO QCD expression for the squared radial wavefunction at the origin in the SIM scheme~\cite{Maxia:2026fbz} is given by
\begin{equation}
\begin{aligned}
&\big\vert R^{\rm SIM}_{n,s}(0)\big\vert^2=\dfrac{1}{2}\left(\dfrac{C_F\alpha_s(\mu_C)m_t}{n}\right)^{\!3}\Bigg[1+\dfrac{\alpha_s(\mu_C)}{4\pi}\left(6\beta_0L_n+c_{\psi,1}\right)\\
&+\left(\dfrac{\alpha_s(\mu_C)}{4\pi}\right)^2\left(24\beta_0^2L_n^2+L_n\left(-12\beta_0^2+6\beta_1+8\beta_0c_{\psi,1}\right)+c_{\psi,2}\right)\\
&+\alpha_s^2(\mu_C) \bigg(C_F^2\left(\frac{2L_n-2S_1(n)}{2s+1}-\frac{15}{8n^2}+\frac{4}{(2s+1)n}\right)+C_FC_A\left(L_n-S_1(n)+\frac{2}{n}\right)\bigg)\Bigg]\\
&\times\Bigg[1+\alpha_s^2(\mu_R)\log{\left(\frac{m_t^2}{\mu_R^2}\right)}\bigg(\frac{C_F^2}{2s+1}+\frac{C_FC_A}{2}\bigg)\Bigg]\\
&\times \exp{\!\left[32\pi^2\left(\frac{C_F^2}{2s+1}+\frac{C_FC_A}{2}\right)\frac{\arctan{\!\left(\frac{\beta_2\frac{\alpha_s(\mu_C)}{2\pi}+\beta_1}{\sqrt{4\beta_0\beta_2-\beta_1^2}}\right)}-\arctan{\!\left(\frac{\beta_2\frac{\alpha_s(\mu_R)}{2\pi}+\beta_1}{\sqrt{4\beta_0\beta_2-\beta_1^2}}\right)}}{\sqrt{4\beta_0\beta_2-\beta_1^2}}\right]},
\end{aligned}
\end{equation}
where $c_{\psi,1}$ and $c_{\psi,2}$ are defined in eqs.~(27) and (28) of ref.~\cite{Beneke:2005hg}, the logarithm is defined as
\begin{equation}
    L_n \equiv \log\!\left(\dfrac{n\mu_C}{C_F\alpha_s(\mu_C)m_t}\right),
\end{equation}
and $\beta_i$ are the coefficients of the QCD beta function in the perturbative expansion in $\alpha_s/(4\pi)$. $C_F=4/3$ and $C_A=3$ are QCD Casimir factors and $S_a(n)=\sum_{k=1}^{n}{k^{-a}}$ is the harmonic sum. The exponential term resums the hard-to-soft logarithms. The NLO QED corrections to the squared radial wavefunction at the origin [cf.\@ eq.~(53) of ref.~\cite{Pineda:2006ri}] are given by
\begin{equation}
\begin{aligned}
\big\vert R^{\rm QED}_{n}(0)\big\vert^2=&\dfrac{1}{2}\left(\dfrac{C_F\alpha_s(\mu_C)m_t}{n}\right)^{\!3}\Bigg\{\frac{3e_t^2\alpha}{C_F\alpha_s(\mu_C)}\\
&+\frac{e_t^2\alpha}{\pi C_F}\left[\frac{3\pi e_t^2\alpha}{C_F\alpha_s^2(\mu_C)}+\frac{3}{2}a_1+\beta_0\left(3L_n+S_1(n)+2nS_2(n)-\frac{5}{2}-2n\zeta_2\right)\right]\Bigg\}\,,
\end{aligned}
\end{equation}
where $e_t=2/3$,
\begin{equation}
a_1\equiv\frac{31}{9}C_A-\frac{20}{9}T_Fn_f
\end{equation}
with $T_F=1/2$ and $n_f=5$, and $\zeta_k$ is the Riemann zeta function $\zeta_k\equiv\zeta(k)$. The LO top-Yukawa correction to the squared radial wavefunction at the origin~\cite{Beneke:2015lwa} is given by
\begin{equation}
|R_n^{\rm Yuk}(0)|^2=\dfrac{1}{2}\left(\dfrac{C_F\alpha_s(\mu_C)m_t}{n}\right)^{\!3}\frac{C_F\alpha_s(\mu_C)}{4\pi}\frac{y_t^2}{2}\frac{m_t^2}{m_H^2}\Bigg[2L_n+1+\frac{4}{n}-2S_1(n)\Bigg]\,,
\end{equation}
where $y_t=\sqrt{2}m_t/v$ is the top-quark Yukawa coupling, with $v=\left(\sqrt{2}G_F\right)^{-\frac{1}{2}}\approx 246\,\mathrm{GeV}$ denoting the Higgs vacuum expectation value. Therefore, the $S$-wave LDMEs at NNLO are given by
\begin{eqnarray}
\langle \mathcal{O}_{\etat}\rangle&=&\sum_{n=1}^{+\infty}{\langle \mathcal{O}_{n,0}^{\rm SIM}\rangle}\,,\\
\langle \mathcal{O}_{\Thetat}\rangle&=&\sum_{n=1}^{+\infty}{\langle \mathcal{O}_{n,1}^{\rm SIM}\rangle}\,.
\end{eqnarray}

For $P$-wave states, only the LO LDME is required:
\begin{equation}
\langle \mathcal{O}_{\chit}\rangle=(2J+1)\frac{2N_c}{\pi}\left(\frac{C_F\alpha_s(\mu_C)m_t}{2}\right)^5\sum_{n=1}^{+\infty}{\frac{n^2-1}{n^5}}=(2J+1)\frac{2N_c}{\pi}\left(\frac{C_F\alpha_s(\mu_C)m_t}{2}\right)^5\left(\zeta_3-\zeta_5\right)\,.
\end{equation}
For the relativistic $v^2$ correction, the LDME obtained from the Gremm-Kapustin relation~\cite{Gremm:1997dq}, as defined in eq.~\eqref{eq:sigmadecom}, is
\begin{equation}
\langle \mathcal{P}_{\eta_{t}}\rangle=-\frac{2N_c}{\pi}\left(\frac{C_F\alpha_s(\mu_C)m_t}{2}\right)^5\sum_{n=1}^{+\infty}{\frac{1}{n^5}}=-\frac{2N_c}{\pi}\left(\frac{C_F\alpha_s(\mu_C)m_t}{2}\right)^5\zeta_5\,.
\end{equation}

\subsection{Cross sections\label{sec:SMtables}}

In this subsection, we present the decomposition of the cross section in table~\ref{tab:xsec_decomposition}, corresponding to the results shown in figure~\ref{fig:xsec_decomposition}. To present the breakdown of relative corrections, we define
\begin{equation}
\sigma^{(v^2+P\text{-wave})}=\hat{\sigma}_{\etat,v^2}\langle\mathcal{P}_{\etat}\rangle
+\hat{\sigma}_{\chitz}\langle\mathcal{O}_{\chitz}\rangle
+\hat{\sigma}_{\chitt}\langle\mathcal{O}_{\chitt}\rangle\,.
\end{equation}

\begin{figure}
    \centering
    \includegraphics[height=8.6cm]{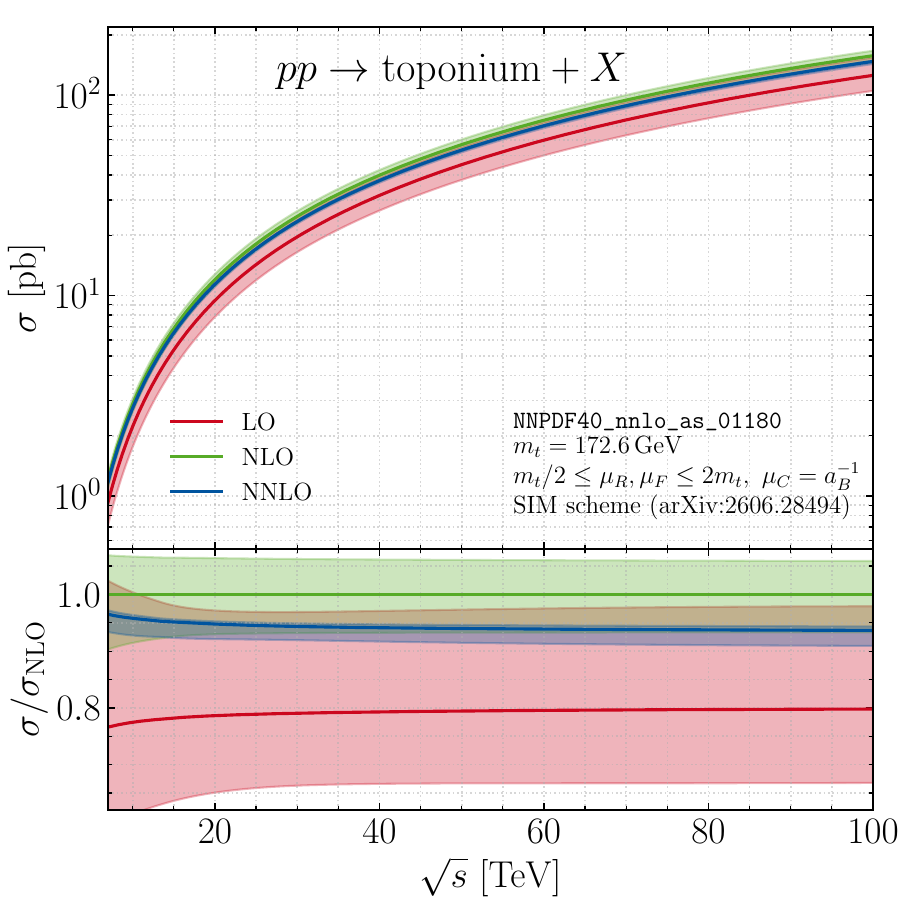}
    \caption{Total inclusive toponium-production cross sections at LO (red), NLO (green), and NNLO (blue) in the upper panel, and the corresponding cross-section ratios normalized to the NLO central values in the lower panel, as functions of $\sqrt{s}$. The shaded bands indicate the scale uncertainties estimated from $7$-point variations of $\mu_R$ and $\mu_F$.}
    \label{fig:xsec_total}
\end{figure}

\begin{figure}
    \centering
    \includegraphics[height=8.6cm]{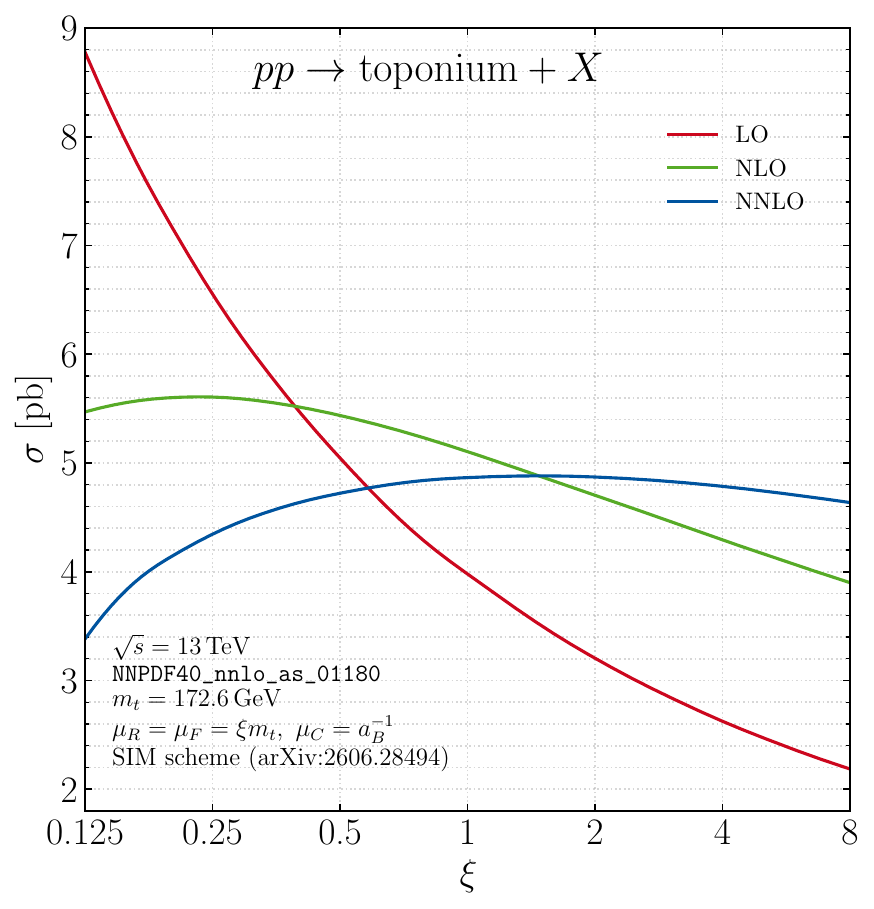}
    \caption{Renormalization- and factorization-scale dependence of the total LO, NLO, and NNLO cross sections for toponium production at the LHC with $\sqrt{s}=13\,\mathrm{TeV}$, including all contributions in our power counting. We set $\mu_R=\mu_F=\xi m_t$.}
    \label{fig:scale_dependence_all}
\end{figure}

Additionally, figure~\ref{fig:xsec_total} shows the total cross sections, including all contributions up to LO (red), NLO (green), and NNLO (blue), with numerical values given in table~\ref{tab:xsec_total}. For $\sqrt{s}=13\,\mathrm{TeV}$, figure~\ref{fig:scale_dependence_all} shows the renormalization- and factorization-scale dependence of the total toponium-production cross sections over the range $1/8<\xi<8$. This complements figure~\ref{fig:scale_dependence}, which shows the corresponding dependence for the dominant channel, $pp\to\etat+X$, including only QCD corrections.

We finally present the LO cross sections for the six associated toponium-production processes in table~\ref{tab:assocxs}, which contribute to the inclusive cross section at NNLO in our power counting. The two leading contributions come from $gg\to\etat H$ and $gg\to \Thetat \gamma$.

%=======================================================================%
% TABLE: Figure 1
%=======================================================================%
\begin{table}[t]
\centering
\renewcommand*{\arraystretch}{1.5}
\begin{tabular}{|c|c|c|c|c|c|c|c|c|c|}
\hline
\multicolumn{2}{|c|}{$\sqrt{s}$ [TeV]}
& $7$ & $8$ & $13$ & $13.6$ & $14$ & $27$ & $85$ & $100$ \\
\hline
$\etat\ \text{NLO QCD}$ & $\hat{\sigma}^{(1)}_{\etat}\langle\mathcal{O}_{\etat}\rangle/\sigma_{\mathrm{LO}}$ [\%] & $29.7$ & $29.1$ & $27.3$ & $27.2$ & $27.1$ & $25.5$ & $24.1$ & $24.0$ \\
$\etat\ \text{NNLO QCD}$ & $\hat{\sigma}^{(2)}_{\etat}\langle\mathcal{O}_{\etat}\rangle/\sigma_{\mathrm{LO}}$ [\%] & $0.4$ & $0.2$ & $-0.6$ & $-0.7$ & $-0.7$ & $-1.4$ & $-2.1$ & $-2.1$ \\
$\etat\ \text{NLO EW}$ & $\hat{\sigma}^{(\mathrm{EW})}_{\etat}\langle\mathcal{O}_{\etat}\rangle/\sigma_{\mathrm{LO}}$ [\%] & $-7.0$ & $-7.0$ & $-7.0$ & $-7.0$ & $-7.0$ & $-7.0$ & $-7.0$ & $-7.0$ \\
$v^2+P\text{-wave}$ & $\sigma^{(v^2+P\text{-wave})}/\sigma_{\mathrm{LO}}$ [\%] & $0.8$ & $0.8$ & $0.8$ & $0.8$ & $0.8$ & $0.8$ & $0.8$ & $0.8$ \\
$q\overline{q}\to\gamma^\ast/Z^\ast\to\Thetat$ & $\hat{\sigma}_{\Thetat}^{(q\bar{q}\to\gamma^*/Z^*\to \Thetat)}\langle\mathcal{O}_{\Thetat}\rangle/\sigma_{\mathrm{LO}}$ [\%] & $1.2$ & $1.1$ & $0.9$ & $0.9$ & $0.9$ & $0.7$ & $0.6$ & $0.6$ \\
$\Thetat\ \text{QCD}$ & $\left(\hat{\sigma}_{\Thetat}^{(1)}+\hat{\sigma}_{\Thetat}^{(2)}\right)\langle\mathcal{O}_{\Thetat}\rangle/\sigma_{\mathrm{LO}}$ [\%] & $0.7$ & $0.7$ & $0.7$ & $0.7$ & $0.7$ & $0.8$ & $0.9$ & $0.9$ \\
$\text{associated}$ & $\left(\hat{\sigma}_{\etat}^{(\mathrm{assoc.})}\langle\mathcal{O}_{\etat}\rangle+\hat{\sigma}_{\Thetat}^{(\mathrm{assoc.})}\langle\mathcal{O}_{\Thetat}\rangle\right)/\sigma_{\mathrm{LO}}$ [\%] & $0.1$ & $0.1$ & $0.1$ & $0.1$ & $0.1$ & $0.2$ & $0.2$ & $0.2$ \\
\hline
\end{tabular}
\caption{Relative corrections to the LO cross section as functions of $\sqrt{s}$, as shown in figure~\ref{fig:xsec_decomposition}.}
\label{tab:xsec_decomposition}
\end{table}

\begin{table}[t]
\centering
\renewcommand*{\arraystretch}{1.5}
\begin{tabular}[t]{|c|c|c|c|c|c|c|c|c|}
\hline
$\sqrt{s}$ [TeV] & $7$ & $8$ & $13$ & $13.6$ & $14$ & $27$ & $85$ & $100$ \\
\hline
%$\sigma_{\etat}^{B}$ & $400^{+157}_{-104}\,\mathrm{ab}$ & $599^{+224}_{-150}\,\mathrm{ab}$ & $2.19^{+0.69}_{-0.49}\,\mathrm{fb}$ & $2.45^{+0.76}_{-0.54}\,\mathrm{fb}$ & $2.62^{+0.81}_{-0.57}\,\mathrm{fb}$ & $10.9^{+2.59}_{-1.97}\,\mathrm{fb}$ & $79.2^{+18.0}_{-13.1}\,\mathrm{fb}$ & $101^{+23}_{-17}\,\mathrm{fb}$ \\
$\sigma_{\etat}^{H}$ & $382^{+149}_{-98}\,\mathrm{ab}$ & $571^{+213}_{-142}\,\mathrm{ab}$ & $2.08^{+0.66}_{-0.46}\,\mathrm{fb}$ & $2.33^{+0.72}_{-0.50}\,\mathrm{fb}$ & $2.49^{+0.77}_{-0.54}\,\mathrm{fb}$ & $10.3^{+2.5}_{-1.8}\,\mathrm{fb}$ & $74.8^{+17.0}_{-12.2}\,\mathrm{fb}$ & $95.7^{+21.7}_{-15.6}\,\mathrm{fb}$ \\
$\sigma_{\etat}^{Z}$ & $18.4^{+7.4}_{-4.9}\,\mathrm{ab}$ & $28.0^{+10.8}_{-7.2}\,\mathrm{ab}$ & $108^{+35}_{-24}\,\mathrm{ab}$ & $121^{+39}_{-27}\,\mathrm{ab}$ & $130^{+42}_{-29}\,\mathrm{ab}$ & $568^{+142}_{-105}\,\mathrm{ab}$ & $4.39^{+1.00}_{-0.71}\,\mathrm{fb}$ & $5.65^{+1.29}_{-0.92}\,\mathrm{fb}$ \\
$\sigma_{\etat}^{\ttbar}$ & $2.85^{+2.43}_{-1.21}\,\mathrm{ab}$ & $5.33^{+4.42}_{-2.23}\,\mathrm{ab}$ & $38.8^{+29.4}_{-15.5}\,\mathrm{ab}$ & $45.8^{+34.4}_{-18.2}\,\mathrm{ab}$ & $50.8^{+37.9}_{-20.1}\,\mathrm{ab}$ & $415^{+275}_{-153}\,\mathrm{ab}$ & $6.83^{+3.66}_{-2.21}\,\mathrm{fb}$ & $9.56^{+4.96}_{-3.03}\,\mathrm{fb}$ \\
%$\sigma_{\Thetat}^{B}$ & $610^{+227}_{-153}\,\mathrm{ab}$ & $894^{+319}_{-217}\,\mathrm{ab}$ & $3.10^{+0.94}_{-0.67}\,\mathrm{fb}$ & $3.45^{+1.03}_{-0.74}\,\mathrm{fb}$ & $3.69^{+1.09}_{-0.78}\,\mathrm{fb}$ & $14.7^{+3.35}_{-2.57}\,\mathrm{fb}$ & $103^{+23}_{-17}\,\mathrm{fb}$ & $132^{+30}_{-22}\,\mathrm{fb}$ \\
$\sigma_{\Thetat}^{\gamma}$ & $579^{+215}_{-144}\,\mathrm{ab}$ & $848^{+302}_{-204}\,\mathrm{ab}$ & $2.93^{+0.88}_{-0.63}\,\mathrm{fb}$ & $3.26^{+0.97}_{-0.69}\,\mathrm{fb}$ & $3.49^{+1.02}_{-0.73}\,\mathrm{fb}$ & $13.8^{+3.1}_{-2.4}\,\mathrm{fb}$ & $96.9^{+21.9}_{-15.9}\,\mathrm{fb}$ & $123^{+28}_{-20}\,\mathrm{fb}$ \\
$\sigma_{\Thetat}^{Z}$ & $30.3^{+11.9}_{-7.8}\,\mathrm{ab}$ & $45.5^{+17.1}_{-11.4}\,\mathrm{ab}$ & $170^{+54}_{-38}\,\mathrm{ab}$ & $190^{+60}_{-42}\,\mathrm{ab}$ & $204^{+64}_{-45}\,\mathrm{ab}$ & $866^{+210}_{-158}\,\mathrm{ab}$ & $6.49^{+1.47}_{-1.05}\,\mathrm{fb}$ & $8.33^{+1.89}_{-1.35}\,\mathrm{fb}$  \\
$\sigma_{\Thetat}^{\ttbar}$ & $6.45^{+4.84}_{-2.56}\,\mathrm{ab}$ & $10.7^{+7.9}_{-4.2}\,\mathrm{ab}$ & $55.1^{+38.4}_{-20.9}\,\mathrm{ab}$ & $63.4^{+43.8}_{-23.9}\,\mathrm{ab}$ & $69.3^{+47.7}_{-26.1}\,\mathrm{ab}$ & $430^{+266}_{-153}\,\mathrm{ab}$ & $5.34^{+2.67}_{-1.66}\,\mathrm{fb}$ & $7.25^{+3.62}_{-2.20}\,\mathrm{fb}$ \\
\hline
\end{tabular}
\caption{Cross sections for associated production processes at LHC and FCC-hh energies, with uncertainties estimated from seven-point renormalization- and factorization-scale variations.}
\label{tab:assocxs}
\end{table}

%TC:endignore
\end{document}